\documentclass[11pt]{article}

\usepackage{graphicx}

\usepackage{mathptmx}

\usepackage{amsfonts, amssymb, latexsym,color, amscd, amsmath}
\usepackage[english]{babel}
\usepackage{graphicx}
\usepackage{fullpage}
\usepackage{cancel}
\usepackage{algorithm,algorithmic}
\usepackage{wasysym}
\usepackage{natbib}

\newenvironment{ex}
   {\paragraph{\bf Example}}
   {\hfill$\blacktriangleleft$ \bigskip}

\newsavebox{\fmbox}

\newcommand{\bu}{\mathbf{u}}

\newcommand{\by}{\mathbf{y}}
\newcommand{\bz}{\mathbf{z}}

\newcommand{\btheta}{\boldsymbol{\theta}}
\newcommand{\feta}{\boldsymbol{\eta}}

\usepackage{geometry}
\usepackage{setspace} 

\begin{document}

\title{Approximate Bayesian Computational methods\thanks{This research was financially supported by the French Agence Nationale de la
Recherche grant 'EMILE' ANR-09-BLAN-0145-01, as well as by the Fondation des Sciences Math\'ematiques de Paris and a GIS
scholarship for the fourth author.}
}

\author{Jean-Michel Marin\footnote{Corresponding author: jean-michel.marin@univ-montp2.fr} \\ 
Institut de Math\'ematiques et Mod\'elisation (I3M), \\ Universit\'e Montpellier 2, France \and
Pierre Pudlo \\ Institut de Math\'ematiques et Mod\'elisation (I3M), \\ Universit\'e Montpellier 2, France \and
Christian P. Robert \\ CEREMADE, Universit\'e Paris Dauphine and \\ CREST, INSEE, Paris \and
Robin J. Ryder \\ CEREMADE, Universit\'e Paris Dauphine and \\ CREST, INSEE, Paris} 


\maketitle

\begin{abstract}
Also known as likelihood-free methods, approximate Bayesian computational (ABC) methods have appeared in the past ten
years as the most satisfactory approach to intractable likelihood problems, first in genetics then in a broader spectrum
of applications. However, these methods suffer to some degree from calibration difficulties that make them rather volatile
in their implementation and thus render them suspicious to the users of more traditional Monte Carlo methods. In this
survey, we study the various improvements and extensions brought on the original ABC algorithm in recent years.

\vspace{0.5cm}\textbf{Keywords:} likelihood-free methods, Bayesian statistics, ABC methodology, DIYABC, Bayesian model choice
\end{abstract}

\section{Introduction}

Conducting a Bayesian analysis in situations where the likelihood function $\ell(\btheta|\by)$ is not available raises a
computational issue. The likelihood may be unavailable for mathematical reasons (it is not available in closed 
from as a function of $\btheta$) or for computational reasons (it is too expensive too calculate).

In some specific settings, the likelihood is expressed as an intractable multidimensional integral 
\begin{equation*}
\ell(\btheta|\by) = \int \ell^\star(\btheta|\by,\bu) \text{d}\bu\,,
\end{equation*}
where $\by\in\mathcal{D}\subseteq \mathbb{R}^n$ is observed, $\bu\in\mathbb{R}^p$ a latent vector and $\btheta\in\mathbb{R}^d$
the parameter of interest. For instance, when facing coalecent models in population genetics \citep[see, e.g.][]{tavare:balding:griffith:donnelly:1997},
$\by$ is the genotypes of the present sample, while $\bu$ stands for their genealogical tree and the genotypes of their ancestors.
In the particular set-up of hierarchical models with partly conjugate
priors, it may be that the corresponding conditional distributions can be simulated and this property leads to a Gibbs
sampler \citep{gelfand:smith:1990}. Such a decomposition is not available in general and
there is no generic way to implement an MCMC algorithm like the Metropolis--Hastings
algorithm \citep[see, e.g.,][]{robert:casella:2004,marin:robert:2007}. 
Typically, the increase in dimension induced by the data augmentation from $\btheta$ to $\bu$ may be such
that the convergence properties of the corresponding MCMC algorithms are too poor for the algorithm to be
considered. 

In others situations, the normalizing constant of the likelihood $Z_{\btheta}$ is unknown
\begin{equation*}
\ell(\btheta|\by) = \ell_1(\btheta|\by)/ Z_{\btheta}\,.
\end{equation*}
This is typically the case of Gibbs random fields used to model the dependency 
within spatially correlated data, with applications in epidemiology
and image analysis, among others (e.g. \cite{rue:held:2005}).
For such models, a solution relying on the simulation of pseudo-samples has been proposed
by \cite{moeller:pettitt:reeves:2006}. However the dependency of this solution on a pseudo-target
distribution makes it difficult to calibrate \citep{cucala:marin:robert:titterington:2006,friel:pettitt:2008}
in general settings. 

Bayesian inference thus faces a large class of settings where the likelihood function is not completely known,
e.g.~$\ell(\btheta|\by) = \ell_1(\btheta|\by)\ell_2(\btheta)$ with $\ell_2$ unknown, and where exact simulation from the
corresponding posterior distribution is impractical or even impossible. Such settings call for practical if cruder
approximations methods. In the past, Laplace approximations \citep{tierney:kadane:1986} and variational Bayes solutions
\citep{jaakkola:jordan:2000} have been advanced for such problems. However, Laplace approximations require some analytic
knowledge of the posterior distribution, while variational Bayes solutions replace the true model with another
pseudo-model which is usually much simpler and thus misses some of the features of the original model.

The ABC methodology, where ABC stands for {\em approximate Bayesian computation}, was mentioned as early as 1984 through a pedagogical
and philosophical argument in \cite{rubin:1984}. It offers an almost automated resolution of the difficulty
with models which are intractable but can be simulated from. It was first proposed in population genetics by
\cite{tavare:balding:griffith:donnelly:1997}, who introduced approximate Bayesian computational methods as a rejection
technique bypassing the computation of the likelihood function via a simulation from the corresponding distribution.
The exact version of the method cannot be implemented but in a very small range of cases. 
\cite{pritchard:seielstad:perez:feldman:1999} produce a generalisation based on an approximation of the target. We
study here the foundations as well as the implementation of the ABC method, with illustrations from time series.

This survey describes the genesis of the ABC approach and its justifications (Section \ref{Genesis}), the
calibration of the method (Section \ref{ACDC}), recent sequential improvements (Section \ref{Scorpion}),
post-proc\-essing of ABC outputs (Section \ref{Pogues}), and the specific application of ABC to model choice
(Section \ref{RoxyMusic}). The illustrations of the ABC methodology are based on the posteriors of the MA$(2)$
and MA$(1)$ models for which the true posterior distribution can be computed; the impact of the ABC
approximation can thus be assessed.  We do not cover the increasingly wide array of applications of ABC here
here; see \cite{csillery:blum:gaggiotti:francois:2010} for a survey of implementations of ABC in genomics and
ecology.  Neither do we address the controversy raised by \cite{templeton:2008,templeton:2010} about the lack
of validity of the ABC approach in statistical testing. Answers to those criticisms are provided in
\cite{clade:2010,csillery:blum:gaggiotti:francois:2010b,berger:fienberg:raftery:robert:2010}, among others.

\section{Genesis of the ABC approach and  justifications}\label{Genesis}

\paragraph{\bf Prehistory}
\cite{rubin:1984} advances a visionary statement that `Bayesian statistics and Monte Carlo methods are ideally
suited to the task of passing many models over one dataset'. Furthermore, he produces in this paper a
description of the first ABC algorithm.  Followed by \cite{tavare:balding:griffith:donnelly:1997}, the original
ABC algorithm is in fact a special case of an accept-reject method \citep[see, e.g.,][]{robert:casella:2004}, where the
parameter $\btheta$ is generated from the prior $\pi(\btheta)$ and the acceptance is conditional on the
corresponding simulation of a sample being `almost' identical to the (true) observed sample, which is denoted
$\by$ throughout this paper. For the original algorithm given below (and solely for this algorithm), we
suppose that $\by$ takes values in a finite or countable set $\mathcal{D}$.

\begin{algorithm}[H]
\caption{Likelihood-free rejection sampler 1\label{algo:ABC1}}
\begin{algorithmic}
\FOR {$i=1$ to $N$}
	\REPEAT
	\STATE Generate $\btheta'$ from the prior distribution $\pi(\cdot)$
        \STATE Generate $\bz$ from the likelihood $f(\cdot|\btheta')$
        \UNTIL {$\bz = \by$}
           \STATE set $\btheta_i=\btheta'$,
\ENDFOR
\end{algorithmic}
\end{algorithm}

\newpage

It is straightforward to show that the outcome $\big(\btheta_1,$ $\btheta_2,\ldots, \btheta_N\big)$ 
resulting from this algorithm is an iid sample from the posterior distribution since
\begin{align*}
f(\btheta_i) &\propto \sum_{\bz\in\mathcal{D}}\pi(\btheta_i)f(\bz|\btheta_i)\mathbb{I}_{\by}(\bz)
             = \pi(\btheta_i)f(\by|\btheta_i) \\
             &\propto \pi(\btheta_i|\by)\,.
\end{align*}
\cite{rubin:1984} does not promote this simulation method in situations where the likelihood is not available but rather exhibits it 
as an intuitive way to understand posterior distributions from a frequentist perspective, because parameters from the
posterior are more likely to be those that could have generated the observed data. (The issue of the zero probability of
the exact equality between simulated and observed data in continuous settings is not addressed in the original paper,
presumably because the very notion of a `match' between simulated and observed data is not precisely defined.)

\paragraph{\bf The first ABC}
In a population genetics setting, \cite{pritchard:seielstad:perez:feldman:1999} extend the above algorithm to the
case of continuous sample spaces, producing the first genuine ABC algorithm, defined as follows

\begin{algorithm}[H]
\caption{Likelihood-free rejection sampler 2\label{algo:ABC2}}
\begin{algorithmic}
\FOR {$i=1$ to $N$}
	\REPEAT
	\STATE Generate $\btheta'$ from the prior distribution $\pi(\cdot)$
        \STATE Generate $\bz$ from the likelihood $f(\cdot|\btheta')$
        \UNTIL {$\rho\{\eta(\bz),\eta(\by)\}\leq \epsilon$}
           \STATE set $\btheta_i=\btheta'$,
\ENDFOR
\end{algorithmic}
\end{algorithm}

\smallskip
\noindent where the parameters of the algorithm are
\begin{itemize}
\item[--] $\eta$,   a function on $\mathcal{D}$ defining a statistic which most often is not sufficient,
\item[--] $\rho>0$,   a distance on $\eta(\mathcal{D})$,
\item[--] $\epsilon>0$,   a tolerance level.
\end{itemize}

The likelihood-free algorithm above thus samples from the marginal in $\bz$ of the joint distribution
\begin{equation}\label{eq:abctarget}
\pi_\epsilon(\btheta,\bz|\by)=
\frac{\pi(\btheta)f(\bz|\btheta)\mathbb{I}_{A_{\epsilon,\by}}(\bz)}
{\int_{A_{\epsilon,\by}\times\btheta}\pi(\btheta)f(\bz|\btheta)\text{d}\bz\text{d}\btheta}\,,
\end{equation}
where $\mathbb{I}_B(\cdot)$ denotes the indicator function of the set $B$ and
\begin{equation*}
A_{\epsilon,\by}=\{\bz\in\mathcal{D}|\rho\{\eta(\bz),\eta(\by)\}\leq \epsilon\} \,.
\end{equation*}
The basic idea behind ABC is that using a representative (enough) summary statistic $\eta$
coupled with a small (enough) tolerance $\epsilon$ should produce a good (enough) approximation to the posterior
distribution, namely that
\begin{equation*}
\pi_\epsilon(\btheta|\by)=\int \pi_\epsilon(\btheta,\bz|\by)\text{d}\bz\approx \pi(\btheta|\by)\,.
\end{equation*}

Before moving to the extensions of the above algorithm, let us consider a simple dynamic example.

\begin{ex}\label{ex:myfirstMAq} 
The MA$(q)$ process is a stochastic process $(y_k)_{k\in\mathbb{N}^*}$ defined by 
\begin{equation}\label{eq:maq}
y_k=u_k +\sum_{i=1}^q \theta_i u_{k-i} \,,
\end{equation}
where $(u_k)_{k\in\mathbb{Z}}$ is an iid sequence of standard Gaussians $\mathcal{N}(0,1)$. 
Even though a Bayesian analysis can handle non-identifiable settings
and still estimate properly identifiable quantities \citep[see, e.g.,][Chapter 5]{marin:robert:2007}, 
we will impose a standard identifiability condition on this model, namely that the roots of the polynomial
\begin{equation*}
\mathcal{Q}(x) = 1 - \sum_{i=1}^q \theta_i x^i
\end{equation*}
are all outside the unit circle in the complex plane. A simple prior distribution
is therefore the uniform distribution over the corresponding range of $\theta_i$'s, especially when $q$ is
small and the set of resulting parameters is easy to describe. In the case processed in the figures below for $q=2$, we obtain 
the triangle
\begin{equation*}
-2<\theta_1<2\,,\quad
\theta_1+\theta_2>-1\,,\quad
\theta_1-\theta_2<1\,.
\end{equation*}

\begin{figure}[hbtp]
\centering
\includegraphics[width=.5\textwidth]{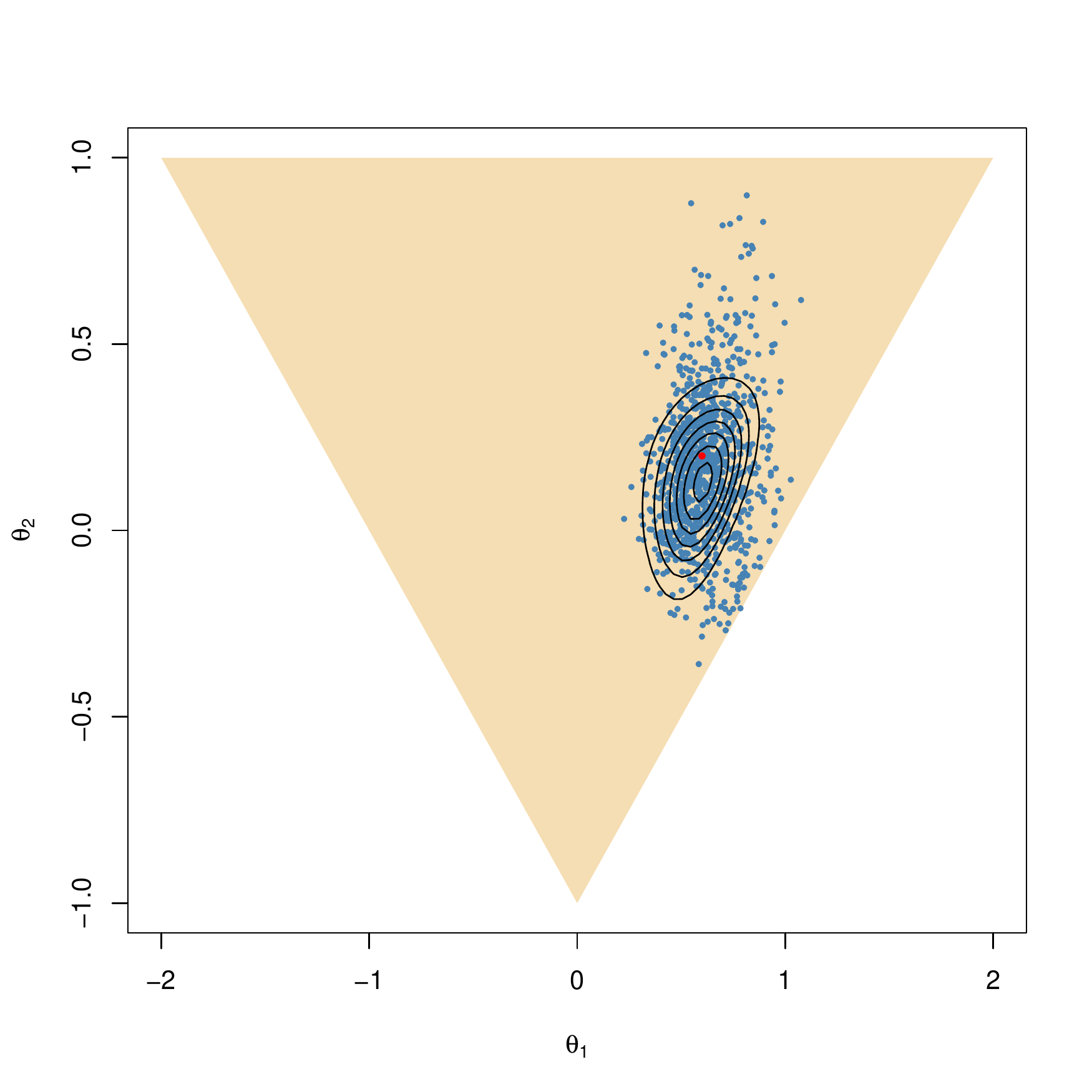}
\caption{\label{fig:basiABc} 
Comparison of the level sets \textit{(in black)} of the true posterior distribution with the scatter plot \textit{(in blue)} 
of an ABC sample when 
using autocovariances as summary statistics. The threshold $\epsilon$ is chosen so that $0.1\%$ of the $N=10^6$ simulated datasets 
are accepted.
The observed dataset has been drawn from an MA$(2)$ model with $n=100$ epochs and parameter $\btheta=(0.6,\,0.2)$ 
\textit{(the red dot)}. The triangle is the range of acceptable values of $\btheta$.
}
\end{figure}

Although the prior on $\btheta$ is very simple, and despite the Gaussian nature of the random variables, the
likelihood associated with a series $(y_k)_{1\le k \le n}$ is more complex because of the need to integrate out
$u_{-q+1},\ldots,u_{-1},u_0$. (The easier alternative is to condition on $(y_k)_{1\le k\le q}$, see
\citealp{marin:robert:2007}, even though the general case can also be handled by MCMC simulations as the
likelihood is available, at least for small values of $n$.) 

Running one iteration of ABC in this setting then simply requires 
\begin{itemize}
\item[(a)] simulating the MA$(q)$ coefficients $\theta$ uniformly over the acceptable range, 
\item[(b)] generating an iid sequence $(u_k)_{-q<k\le n}$,
\item[(c)] producing a simulated series $(z_k)_{1\le k\le n}$.
\end{itemize}
Depending on the focus of the
analysis, the distance can be the raw distance between the series
\begin{equation*}
\rho^2\{(z_k)_{1\le k\le n},(y_k)_{1\le k\le n}\} = \sum_{k=1}^n (y_k-z_k)^2
\end{equation*}
or the quadratic distance between summary statistics like the first $q$ autocovariances
\begin{equation*}
\tau_j = \sum_{k=j+1}^n y_ky_{k-j}
\end{equation*}
which is our choice for the illustration provided in Figure \ref{fig:basiABc}. This experiment shows how an ABC
sample fits the level sets of the true posterior density for a simulated sample of length 
$100$ using the parameters $(\theta_1,\theta_2)=(0.6, 0.2)$ and a tolerance level equal to the $0.1\%$ quantile 
of the sample of the distances. (The level sets were computed from the exact likelihood for the MA$(2)$ model
and a grid of values of $\btheta$ over the acceptable range.) This plot illustrates how the distribution of the sample points  
departs from true posterior:  the approximation 
does not reconstruct  the posterior perfectly. Decreasing $\epsilon$ would lead to a better concentration   of the posterior density on the
level sets, but at the expense of the size of the resulting sample or at a higher computing cost.
\end{ex}

\paragraph{\bf MCMC-ABC}
In practice, using simulations from the prior distribution $\pi(\cdot)$ is inefficient because this does not
account for the data at the proposal stage and thus leads to proposed values located in low posterior probability
regions. As an answer to this problem, \cite{marjoram:etal:2003}  introduce an MCMC-ABC algorithm 
(Algorithm \ref{algo:ABCMCMC}) targeting the approximate posterior distribution $\pi_\epsilon$ of equation
\eqref{eq:abctarget}.

\begin{algorithm}[H]
\caption{Likelihood-free MCMC sampler\label{algo:ABCMCMC}}
\begin{algorithmic}
\STATE Use Algorithm \ref{algo:ABC2} to get a realisation $(\btheta^{(0)},\bz^{(0)})$ 
from the ABC target distribution $\pi_\epsilon(\btheta,\bz|\by)$
\FOR {$t=1$ to $N$} 
\STATE Generate $\btheta'$ from the Markov kernel $q\left(\cdot|\btheta^{(t-1)}\right)$,
\STATE Generate $\bz'$ from the likelihood $f(\cdot|\btheta')$,
\STATE Generate $u$ from $\mathcal{U}_{[0,1]}$,
\IF {$u \leq \dfrac{\pi(\btheta')q(\btheta^{(t-1)}|\btheta')}{\pi(\btheta^{(t-1)})q(\btheta'|\btheta^{(t-1)})}$ and $\rho\{\eta(\bz'),\eta(\by)\}\leq \epsilon$}
\STATE set $(\btheta^{(t)},\bz^{(t)})=(\btheta',\bz')$ 
\ELSE 
\STATE $(\btheta^{(t)},\bz^{(t)})=(\btheta^{(t-1)},\bz^{(t-1)})$,
\ENDIF
\ENDFOR
\end{algorithmic}
\end{algorithm}

\smallskip The acceptance probability used in Algorithm \ref{algo:ABCMCMC} does not involve the calculation of
the likelihood and it thus satisfies ABC requirements. It also produces an MCMC algorithm which exactly targets
$\pi_\epsilon(\btheta,\bz|\by)$ as its stationary distribution. Indeed,
\begin{equation*}
\begin{split}
&\dfrac{\pi_\epsilon(\btheta',\bz'|\by)}{\pi_\epsilon(\btheta^{(t-1)},\bz^{(t-1)}|\by)}
\times \dfrac{q(\btheta^{(t-1)}|\btheta')f(\bz^{(t-1)}|\btheta^{(t-1)})}{q(\btheta'|\btheta^{(t-1)})f(\bz'|\btheta')} \\
&=\dfrac{\pi(\btheta')\,\,{f(\bz'|\btheta')}\,\,\mathbb{I}_{A_{\epsilon,\by}}(\bz')}{\pi(\btheta^{(t-1)})\,\,{f(\bz^{(t-1)}|\btheta^{(t-1)})}
{\mathbb{I}_{A_{\epsilon,\by}}(\bz^{(t-1)})}} \\
&\quad\times \dfrac{q(\btheta^{(t-1)}|\btheta')\,\,{f(\bz^{(t-1)}|\btheta^{(t-1)})}}{q(\btheta'|\btheta^{(t-1)})\,\,{f(\bz'|\btheta')}}\\
&=\dfrac{\pi(\btheta')q(\btheta^{(t-1)}|\btheta')}{\pi(\btheta^{(t-1)}q(\btheta'|\btheta^{(t-1)})}\mathbb{I}_{A_{\epsilon,\by}}(\bz')\,.
\end{split}
\end{equation*}


The initialisation of the MCMC sampler with the rejection sampler (Algorithm~2) can be bypassed since the Markov chain 
forgets its initial state. The computational cost of the initialisation is then reduced. But then we have to run the MCMC
longer to achieve convergence and omit the burn-in first iterations from the output,  which also has a computational cost.
As noted above, the ABC approximation depends on tuning parameters (the summary statistic $\eta$, the
tolerance $\epsilon$, and the distance $\rho$) that have to be chosen prior to running the algorithm and the
calibration of which is discussed in most of the literature. The tolerance $\epsilon$ is somewhat the easiest
aspect of this calibration issue in that, when $\epsilon$ goes to zero, the ABC algorithm becomes exact. 


\paragraph{\bf Noisy ABC}
\cite{wilkinson:2008} proposes to switch perspective, replacing the approximation error resulting from the
loose acceptance condition in the above likelihood-free samplers with an exact inference from a controlled
approximation of the target, essentially a convolution of the regular target with an arbitrary kernel function.
The corresponding ABC target is thus 
\begin{equation}\label{eq:abctarget2} 
\pi_\epsilon(\btheta,\bz|\by)=
\frac{\pi(\btheta)f(\bz|\btheta) K_\epsilon (\by-\bz)} {\int \pi(\btheta)f(\bz|\btheta)K_\epsilon (\by-\bz)
\text{d}\bz\text{d}\btheta}\,, 
\end{equation} 
where $K_\epsilon$ is a well-chosen kernel parameterised by the bandwidth $\epsilon$.  This perspective is
interesting in that the outcome is completely controlled, due to the degree of freedom brought by the choice of
the kernel. \cite{wilkinson:2008} makes the valuable point that if the model includes an error term, then
taking the distribution of that error term to be $K_\epsilon$ leads to an ABC algorithm which simulates exactly
from the error-in-variables posterior. In practice, Wilkinson's (\citeyear{wilkinson:2008}) approach requires a
modification of the standard ABC algorithms, taking into account the kernel $K_\epsilon$ for the simulation of
$\bz$. The new algorithm which includes an accept-reject step imposes an upper bound on the convolution kernel
$K_\epsilon$. 

This perspective of the ``noisy ABC" is also adopted by \cite{fearnhead:prangle:2010} who study the convergence
of ABC based inference. They show that the convolution induced by the kernel representation leads to the true
parameter being the maximum of the integrated log-likelihood and thus that a Bayes estimator is converging to
the true value when the number of observations goes to infinity and the tolerance level goes to zero. They also
stress the connection with the econometrics approach of indirect inference
\cite{gourieroux:monfort:renault:1993}.

\paragraph{\bf ABC Filtering}
\citet{jasra:singh:martin:mccoy:2011} propose an ABC scheme for filtering when the distribution of the
observables conditioned on the hidden state is not available point-wise, related to the convolution particle
filter of \citet{campillo:rossi:2009}.  It is particularly appealing in that it allows complex (hence
realistic) statistical models for filtering.  Theoretical arguments are given to prove that the ABC
approximation of the filter does not accumulate errors along the sequence of observables, when the model has
good mixing properties. \citet{dean:singh:jasra:peters:2011} illustrate this implementation in the specific case of
hidden Markov (HMM) models, relating the ABC implementation with Wilkinson's (\citeyear{wilkinson:2008})
perspective and demonstrating that the pseudo (or noisy) model for which ABC is exact also is an HMM. Using
this representation, they further establish ABC consistency. While \citet{dean:singh:jasra:peters:2011} establish
that ABC leads to an asymptotic bias for a fixed value of the tolerance $\epsilon$, they also prove that an
arbitrary accuracy can be attained with enough data and a small enough $\epsilon$. (We note that the
restriction to summary statistics that preserve the HMM structure is paramount for the results in the paper to
apply, hence preventing the use of truly summarising statistics that would not grow in dimension with the size
of the HMM series.) The convergence result central to \citet{dean:singh:jasra:peters:2011} is also connected with
Fearnhead and Prangle's (2010) version, mentioned above, in that they both rely on pseudo-likelihood
consistency arguments. 

\section{Calibration of ABC}\label{ACDC}

\paragraph{\bf Summary statistics}
Several authors have considered the fundamental difficulty associated with the choice of the summary statistic,
$\eta(\by)$, which one would like to consider as a quasi-sufficient statistic. First, for most real problems (a notable
exception being found in \citealp{grelaud:marin:robert:rodolphe:tally:2009} in the case of Gibbs random fields), it is
impossible to find non-trivial sufficient statistics which would eliminate the need of a choice of statistics. Second,
the summary statistics of interest are usually determined by the problem at hand and chosen by the experimenters in the
field. 

Assuming a large collection of summary statistics is available, \cite{joyce:marjoram:2008} consider the
sequential inclusion of those statistics into the ABC target. The inclusion of a new statistic within the set of summary
statistics is assessed in terms of a likelihood ratio test, without taking into account the sequential nature of the
tests. We have reservations about the method, first and foremost that the construction of the statistics is not
discussed, while the method is not independent from parametrisation, and also that the order in which the statistics
are considered is paramount for their inclusion/exclusion. A regularisation of the method proposed at the end of the
paper is to use a forward-backward selection mechanism to address this last issue. However, this correction does not
address another issue, namely the impact of the correlation between the summary statistics.  Note at last that
Joyce and Marjoram's (2008) method still depends on an approximation factor that needs to be calibrated prior
to running the algorithm. In his thesis, \cite{ratmann:2009} proposes a similar examination of the successive
inclusion of various statistics.

A related perspective is that of \cite{mckinley:cook:deardon:2009}. They perform a simulation experiment comparing
ABC-MCMC and ABC-SMC (discussed below) with regular data augmentation MCMC.  The authors test strategies to
select the tolerance level, and to choose the distance $\rho$ and  the summary statistics.
The conclusions are not very surprising, in that 
\begin{itemize}
\item[(a)] repeating simulations of the data points given one simulated parameter does not seem to contribute to an improved
approximation of the posterior by the ABC sample, 
\item[(b)] the tolerance level does not seem to have a strong influence, 
\item[(c)] the choice of the distance, of the summary statistics and of the calibration factors are
paramount to the success of the approximation, and 
\item[(d)] ABC-SMC outperforms ABC-MCMC (MCMC remaining the reference). 
\end{itemize}

\cite{fearnhead:prangle:2010} study the selection of summary statistics with the interesting perspective that
ABC is then considered from a purely inferential viewpoint and calibrated for estimation purposes. (This
contrasts with most alternative perspectives that envision ABC as a poor man's non-parametric estimation of the
posterior distribution.) \cite{fearnhead:prangle:2010} rely on a randomised version of the summary statistics
from which they derive a well-calibrated version of ABC, \textit{i.e.} an algorithm that gives proper
predictions of given quantities.  The authors consider choices of  summary statistics, and establish that  the
posterior expectations of the parameters of interest are optimal summary statistics, although this follows from
their choice of loss function.

\paragraph{\bf Tolerance threshold and ABC approximation error}
As noted above, the choice of the tolerance level $\epsilon$ is mostly a matter of computational power: smaller
$\epsilon$'s are associated with higher computational costs and the standard practice
\citep{beaumont:zhang:balding:2002} is to select $\epsilon$ as a small percentile of the simulated distances
\begin{equation*}
\rho\{\eta(\bz),\eta(\by)\}.
\end{equation*}
An alternative described below is to set the ABC algorithm within the non-parametric setting of density
estimation, in which case $\epsilon$ is understood as a bandwidth and can be derived from the simulated
population. As noted in \cite{fearnhead:prangle:2010}, this perspective implies that the optimal $\epsilon$ is
then different from zero.

Standing rather apart from other contributions to the field, \cite{ratmann:andrieu:wiujf:richardson:2009} provide an
intrinsically novel way of looking at the ABC approximation error (and hence at the tolerance). It is presented as a tool 
assessing the goodness of fit of a given model. The fundamental idea there is to use the tolerance $\epsilon$ as an
additional parameter of the model, simulating from a joint posterior distribution 
\begin{equation*}
f(\btheta,\epsilon|\by) \propto \xi(\epsilon|\by,\btheta)\pi_{\btheta}(\btheta)\pi_\epsilon(\epsilon)\,, 
\end{equation*}
where $\xi(\epsilon|\by,\btheta)$ plays the role of the likelihood, and $\pi_{\btheta}$ and 
$\pi_\epsilon$ are the corresponding priors on
$\btheta$ and $\epsilon$. In this approach, $\xi(\epsilon|\by,\btheta)$ is the prior predictive density of
$\rho\{\eta(\bz),\eta(\by)\}$ given $\btheta$ and $\by$ when $\bz$ is distributed from $f(\bz|\btheta)$. We note here a
connection with Wilkinson's (\citeyear{wilkinson:2008}) target \eqref{eq:abctarget2} in that $\pi(\btheta)f(\bz|\btheta)
K_\epsilon (\by-\bz)$ is identical to the above once we replace $\by-\bz$ by $\epsilon$.

\cite{ratmann:andrieu:wiujf:richardson:2009} then derive an ABC algorithm they call ABC${}_\mu$ to simulate an MCMC
chain targeting this joint distribution, replacing $\xi(\epsilon|\by,\btheta)$ with a non-parametric kernel
approximation. For each model under comparison, the marginal posterior distribution on the error $\epsilon$ is then used
to assess the fit of the model, the logic being that this posterior should include $0$ in a reasonable credible
interval. While the authors stress they use the data once, they also define the above target by using simultaneously
a prior distribution on $\epsilon$ and a conditional distribution on the same $\epsilon$ that they interpret as the likelihood in
$(\epsilon,\btheta)$. The product is most often defined as a density in $(\epsilon,\btheta)$, so it can be simulated
from, but the Bayesian interpretation of the outcome is delicate, especially because it seems the prior on $\epsilon$
 contributes significantly to the final assessment of the model.  As discussed in \cite{robert:mengersen:chen:2010}, some
of the choices of \cite{ratmann:andrieu:wiujf:richardson:2009} can be argued about, in particular the ambivalent role of
the approximation error. The most important aspect of the paper is that the original motivation of
running ABC for conducting inference on the parameters of a model is replaced by the alternative goal of running ABC for
assessing a model; see Ratmann et al's \citeyear{ratmann:andrieu:wiujf:richardson:2010}  reply to the remarks made by \citet{robert:mengersen:chen:2010}.
.

\begin{figure}[hbtp]
\centering
\includegraphics[width=.5\textwidth]{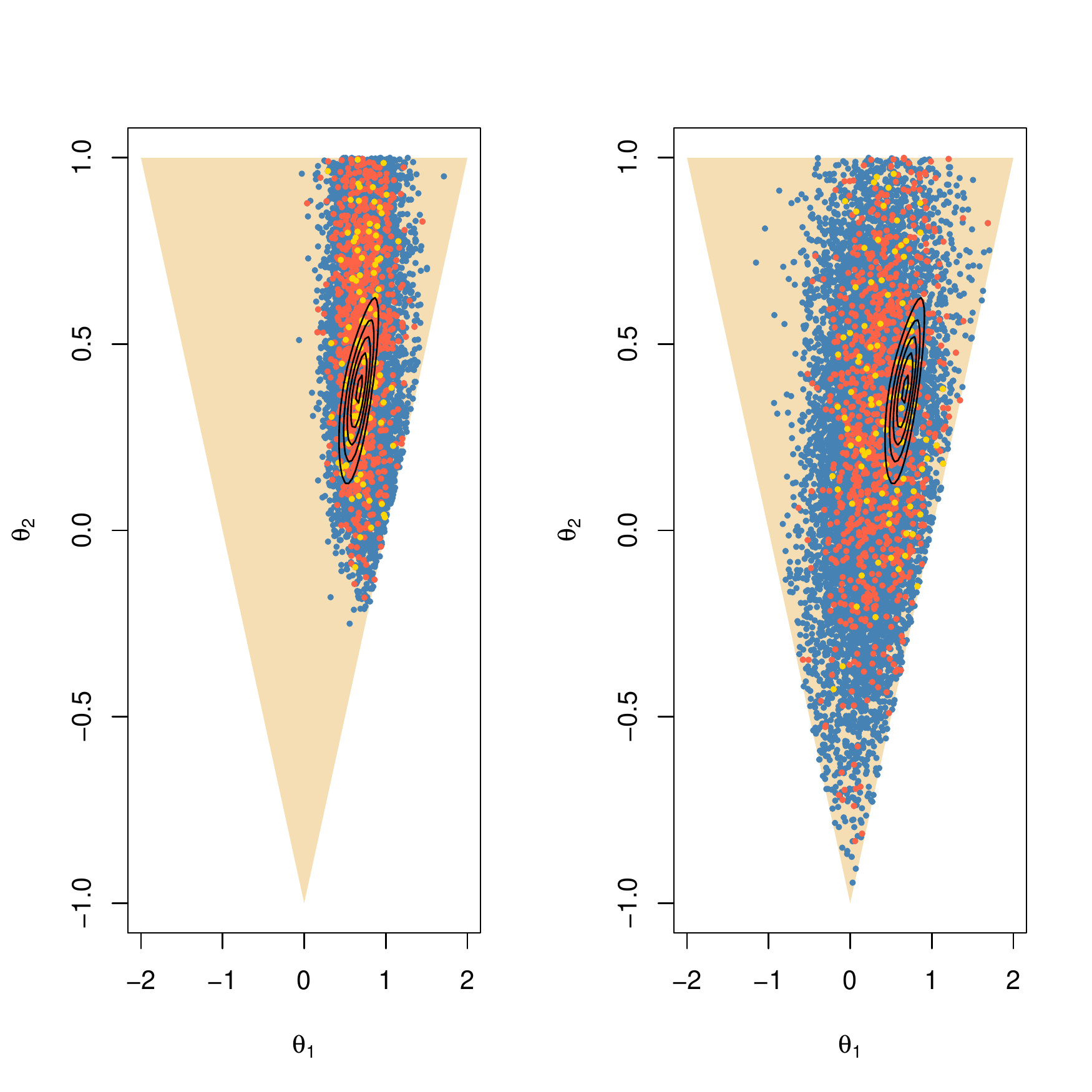}
\caption{\label{fig:autocodist} 
Scattering of two ABC samples when the computations are based on the autocovariance distance 
{\em (left)} and the raw distance {\em (right)}, using different quantiles on the simulated distance for $\epsilon$ 
($1\%$ \textit{in blue}, $1\permil$ \textit{in red}, and $0.1\permil$ \textit{in yellow}). 
The level sets of the posterior density are exhibited \textit{in black}.
}
\end{figure}
\begin{figure}[hbtp]
\centering
\includegraphics[width=.5\textwidth]{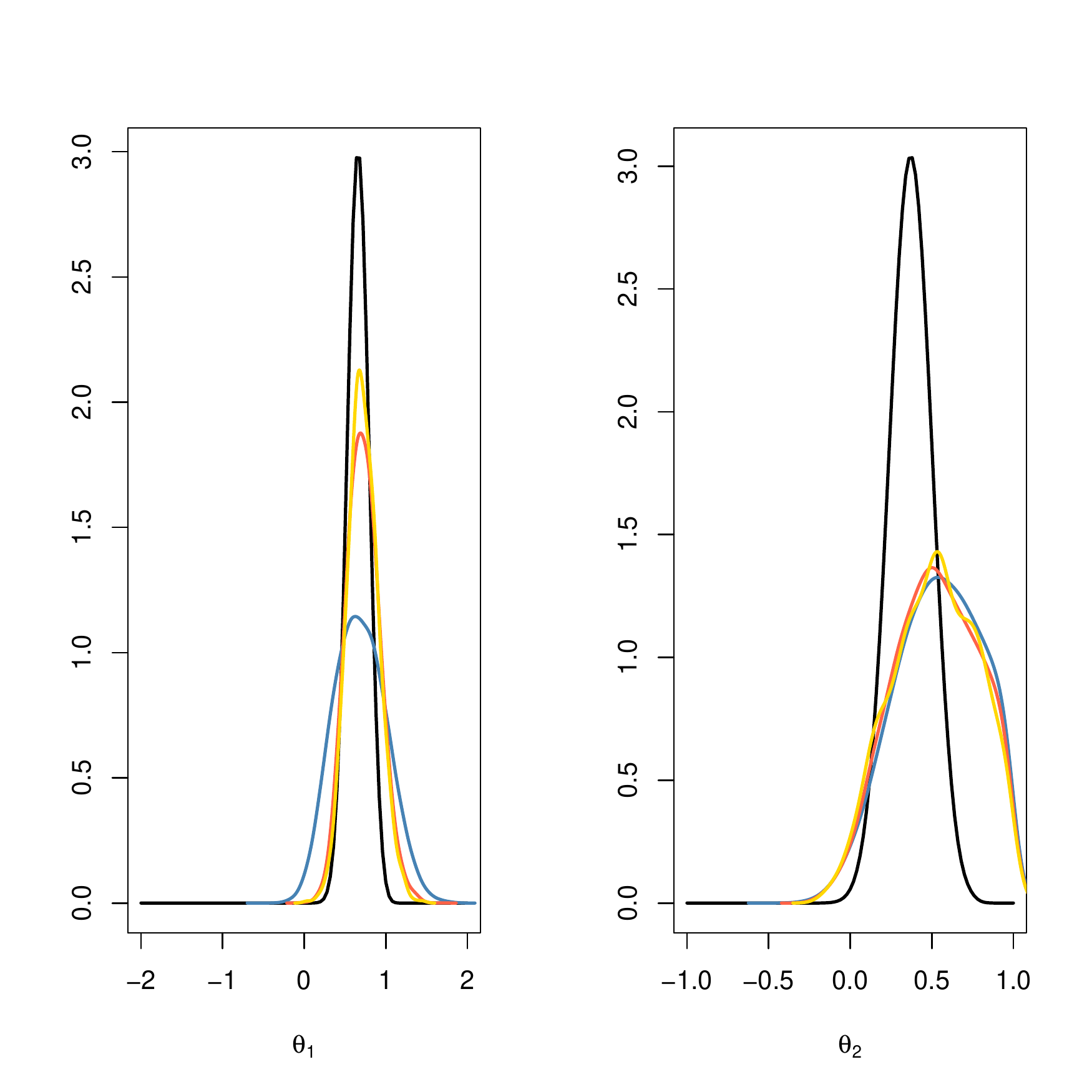}
\caption{\label{fig:rawdist}Evolution of the distribution of ABC samples using 
different quantiles for $\epsilon$ ($10\%$ \textit{in blue}, $1\%$ \textit{in red}, and $0.1\%$ \textit{in yellow}) 
when compared with the true marginal densities. The dataset is the same as in Figure \ref{fig:autocodist}.}
\end{figure}

\begin{ex}
Returning to the MA$(2)$ model, we study the impact of the choice of the distance
and of the tolerance on the approximation. In this example, we simulated a sample of size $50$ from a MA$(2)$ model
based on the same parameters as above. First, we compare the impact of using the raw distance between the complete datasets instead of the distance between the
autocovariances (introduced above). Figure \ref{fig:autocodist} shows that the raw 
distance between the observed and the simulated time series is inefficient and fairly non-discriminative.
For the raw distance, the spread of the parameters accepted after the ABC step is indeed much wider than for the second distance, 
especially when compared with the level sets of the posterior density. We thus use only the distance between the 
autocovariances in the remainder of the paper.

We now turn  to the tolerance $\epsilon$. Figure \ref{fig:rawdist} shows that
decreasing $\epsilon$ along empirical quantiles of the simulated distances $\rho(\eta(\bz),\eta(\by))$ 
improves the approximation, although we never reach the true marginal densities (this is particularly
true for the parameter $\theta_2$.) The marginal densities of the ABC samples were obtained by the {\sf R} default density estimator
and the true marginal densities by numerical integration.
\end{ex}

\section{Sequential improvements}\label{Scorpion}

\paragraph{\bf Importance sampling}
Sequential techniques can enhance the efficiency of the ABC algorithm by learning about the target distribution, as in
\citeauthor{sisson:fan:tanaka:2007}'s (2007) partial rejection control (PRC) version. The ABC-PRC modification introduced 
by \cite{sisson:fan:tanaka:2007} consists in producing samples
$(\btheta_1^{(t)}, \ldots,\btheta_N^{(t)})$ at each iteration $1\le t\le T$ of the algorithm by using a particle filter
methodology. Starting with a regular ABC step, the generation of the $\btheta_i^{(t)}$'s relies on Markov transition kernels $K_t$,
\begin{equation*}
\btheta_i^{(t)} \sim K_t(\btheta|\btheta^\star)\,,
\end{equation*}
until $\bz\sim f(\bz|\btheta_i^{(t)})$ is such that $\rho(\eta(\bz),\eta(\by))\leq\epsilon$,
where $\btheta^\star$ is selected at random among the previous $\btheta_i^{(t-1)}$'s with probabilities
$\omega_i^{(t-1)}$.  The probability $\omega_i^{(t)}$ is derived by an importance sampling argument as
\begin{equation*}
\omega_i^{(t)} \propto \frac{\pi(\btheta_i^{(t)}) L_{t-1}(\btheta^\star|\btheta_i^{(t)})}{
\pi(\btheta^\star)K_t(\btheta_i^{(t)}|\btheta^\star)}\,,\
\end{equation*}
where $L_{t-1}$ is an arbitrary transition kernel. While this method is based upon the theoretical work of
\cite{delmoral:doucet:jasra:2006} and their SMC sampler, the application to approximate Bayesian computation results in
a bias in the approximation to the posterior, because the likelihood is removed in a standard ABC fashion
\citep{sisson:fan:tanaka:2009}. Replacing the likelihood with the indicator function provides an unbiased estimator of
the likelihood that cannot be used as such in the denominator of a Metropolis--Hastings acceptance probability, hence
the resulting bias. 

An alternative version called ABC-PMC and based on genuine importance sampling arguments, proposed by
\cite{beaumont:cornuet:marin:robert:2009}, bypasses this difficulty, in connection with the population Monte
Carlo method of \cite{douc:guillin:marin:robert:2005}. It includes an automatic scaling of the forward kernel. The
correction published in \cite{sisson:fan:tanaka:2009} acknowledges the existence of a bias and suggests a correction
essentially identical to the PMC solution of \cite{beaumont:cornuet:marin:robert:2009}.

As illustrated in the pseudo-code below, ABC-PMC constructs a kernel approximation to the target distribution based on 
earlier simulations and estimates the random walk scale (which is also the kernel bandwidth) from those simulations, using
in addition a decreasing sequence of tolerance thresholds $\epsilon_1\ge\ldots\ge\epsilon_T$:

\begin{algorithm}[H]
\caption{Likelihood-free population Monte Carlo sampler}\label{algo:ABCMC}
\begin{algorithmic}
\STATE At iteration $t=1$,
  \FOR {$i=1$ to $N$}
    \REPEAT
       \STATE Simulate $\btheta_i^{(1)}\sim \pi(\btheta)$ and $\bz\sim f(\bz\mid \btheta_i^{(1)})$ 
     \UNTIL $\rho(\eta(\bz),\eta(\by))\leq\epsilon_1$
    \STATE Set $\omega^{(1)}_i=1/N$
    \ENDFOR
\STATE Take $\Sigma_1$ as twice the empirical variance of the $\btheta_i^{(1)}$'s
\FOR {$t=2$ to $T$}
  \FOR {$i=1$ to $N$}
    \REPEAT
      \STATE Pick $\btheta_i^\star$ from the $\btheta_j^{(t-1)}$'s with probabilities $\omega_j^{(t-1)}$
      \STATE Generate $\btheta_i^{(t)} \sim \mathcal{N}(\btheta_i^\star,\Sigma_{t-1})$ 
	and $\bz\sim f(\bz\mid \btheta_i^{(t)})$
    \UNTIL $\rho(\eta(\bz),\eta(\by))\leq\epsilon_t$\\
    \STATE Set $\omega^{(t)}_i\propto \pi(\btheta^{(t)}_i)/\sum_{j=1}^N \omega^{(t-1)}_j
	\varphi\left\{ \Sigma_{t-1}^{-1/2}\left(\btheta^{(t)}_i-\btheta^{(t-1)}_j\right)\right\}$
     \ENDFOR
\STATE Take $\Sigma_t$ as twice the weighted variance of the $\btheta_i^{(t)}$'s
\ENDFOR
\end{algorithmic}
\end{algorithm}

Another related paper is Toni et al's \citeyear{toni:etal:2009} proposal of a parallel sequential ABC algorithm. Just like ABC-PMC,
the ABC-SMC algorithm (an acronym found in several papers) developed therein is based on a sequence of simulated
samples, Markov transition kernels, and importance weights rather than SMC justifications. The unavailable
likelihood is estimated by the indicator of the tolerance zone or an average of indicators as in
\cite{marjoram:etal:2003}. The bulk of the paper is dedicated to the analysis of ODEs, using uniform distributions as transition
kernels. The adaptivity of the ABC-SMC algorithm is restricted to a progressive reduction of the tolerance,
$\epsilon_t$, since the kernels $K_t$'s remain the same across iterations, in contrast with the ABC-PMC motivation for
tuning the $K_t$'s to the target.
The paper also contains a comparison with ABC-PRC, which shows a bias in the variance of the ABC-PRC output,
in line with \cite{beaumont:cornuet:marin:robert:2009}.

\citet{mckinley:cook:deardon:2009} have coded the parallel sequential ABC algorithm on an infectious disease model 
(a recent outbreak of Ebola Haemorrhagic Fever in the Democratic Republic of the Congo ---
for which there is no known treatment and which is responsible for an $88\%$ decline in observed chimpanzee populations since 2003!).
They show that the ABC-SMC sampler outperforms ABC-MCMC (MCMC remaining the reference).
The comparison experiment is based on a single dataset, with fixed random walk variances for the MCMC algorithms; 
note that the prior used in the simulation might be too highly peaked around the true value (gamma rates of $0.1$). 
Some of the ABC scenarios do produce estimates that are rather far away from the references given by MCMC, 
 for instance CABC-MCMC when the threshold $\epsilon$ is $10$ and the number of repeats $R$ is $100$.

\paragraph{\bf Backward kernels and SMC}
\cite{delmoral:doucet:jasra:2009} exhibit the connection between the ABC algorithm and the foundational SMC paper of
\cite{delmoral:doucet:jasra:2006} that inspired \cite{sisson:fan:tanaka:2007}. As opposed to the latter, and despite a
common framework, this ABC-SMC paper properly relies on the idea of using backward kernels $L_t$ to simplify the
importance weights and to remove  from these weights the dependence on the unknown likelihood. A major assumption of
\cite{delmoral:doucet:jasra:2009} is that the forward kernels $K_t$ are supposed to be invariant against the true target
(which is a tempered-like version of the true posterior in sequential Monte Carlo), a choice not explicitely made
in \cite{sisson:fan:tanaka:2007}.
One of the novelties in the paper is that the authors rely on $M$ repeated simulations of the pseudo-data $\bz$ given
the parameter, rather than using a single simulation. In that perspective, each simulated parameter gets a non-zero
weight that is proportional to the number of accepted $\bz$'s. The limiting case $M\to\infty$ brings in an exact simulation
from the tempered targets $\pi_{\epsilon_t}$'s, so there is a convergence principle and the stabilisation of the approximation could be
assessed to calibrate $M$. The adaptivity in the ABC-SMC algorithm is found in the on-line construction of the
thresholds: the thresholds decrease slowly enough to keep a large number of accepted transitions from the
previous sample.  An important feature is that the update in the importance weights simplifies to the ratio of the
proportions of surviving particles, due to the choice of the reversal backward kernels $L_t$ and to the use of
invariant transition forward kernels $K_t$.

In a very related manner, \cite{drovandi:pettitt:2010} use a combination of particles and of MCMC moves to adapt a
proposal to the true target, with  acceptance probability
\begin{equation*}
\min\left\{1,\dfrac{\pi(\btheta^*)K(\btheta_c|\btheta^*)}{\pi(\btheta^*)K(\btheta^*|\btheta_c)}\right\}
\end{equation*}
where $\btheta^*$ is
the proposed value, $\btheta_c$ is the current value (picked at random from the particle population), and $K$ is a
proposal kernel used to simulate the proposed value. The algorithm is adaptive in that the previous population of
particles is used to make the choice of the proposal $K$, as well as of the tolerance level $\epsilon_t$.
The level of novelty of the method compared with \cite{delmoral:doucet:jasra:2009} is quite limited, since the paper 
adapts the tolerance on-line as an $\alpha$-quantile of the previous particle population. 
The convergence analysis which is omitted by \cite{drovandi:pettitt:2010} is perhaps not so standard, mainly because 
the MCMC is applied only to half of the particle system. \citet{delmoral:doucet:jasra:2011}  tackle the issue of adaptive 
resampling strategies.
The only strong methodological difference between
the two papers is that the MCMC steps are now repeated `numerous times'. However, this  partly cancels the appeal of an
$\text{O}(N)$ order method versus the $\text{O}(N^2)$ order ABC-PMC and ABC-SMC methods. An interesting remark there
is that advances are needed in cases when simulating the pseudo-observations is very costly, as in
Ising models. However, replacing exact simulation by a few steps from a Gibbs sampler 
as in \cite{grelaud:marin:robert:rodolphe:tally:2009} cannot be
very detrimental to the convergence of an approximate algorithm.

\section{Post-processing of ABC output}\label{Pogues}

\paragraph{\bf Local linear regression}
Improvements to the general ABC scheme have been achieved by viewing the problem as a conditional density estimation and
developing techniques to allow for larger $\epsilon$ \citep{beaumont:zhang:balding:2002}. This is a post-processing
scheme in that the simulation process {\em per se} does not change but the analysis of the ABC output does. The authors 
endeavour to include all simulated summary statistics, even those far away from the observed summary statistic, by shrinking the
corresponding parameters in a linear manner. More specifically, they replace the simulated $\btheta$'s with
\begin{equation*}
\btheta^* = \btheta - \{\eta(\bz)-\eta(\by)\}^\text{T}\hat\beta\,,
\end{equation*}
where $\hat\beta$ is obtained by a weighted least squares regression of $\btheta$ on $(\eta(\bz)-\eta(\by))$, using weights of the form
\begin{equation*}
K_\delta\left\{ \rho\{\eta(\bz),\eta(\by)\} \right\}\,,
\end{equation*}
where $K_\delta$ is a non-parametric kernel with bandwidth $\delta$.

\begin{figure}[hbtp]
\centering\includegraphics[width=.5\textwidth]{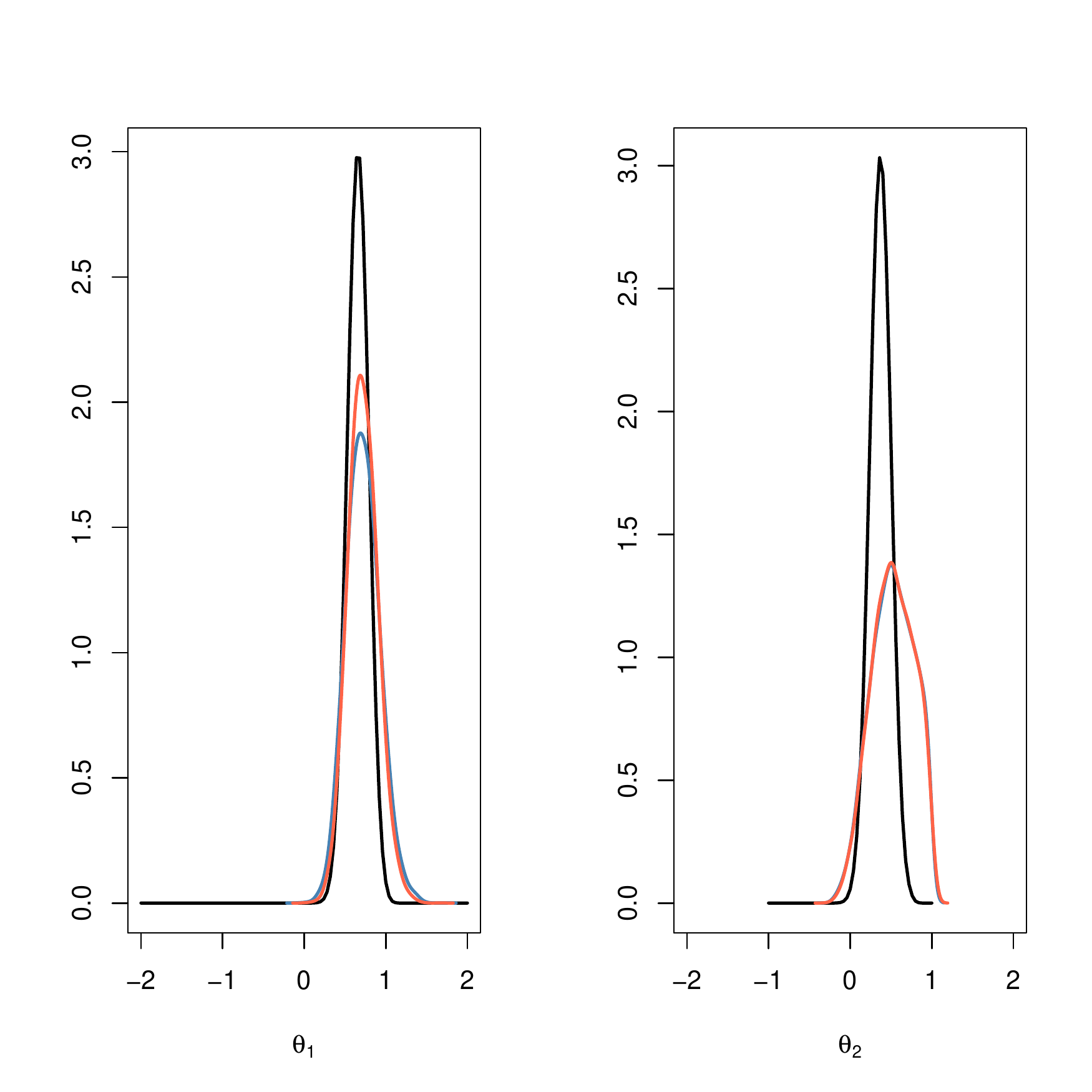}
\caption{\label{fig:BeAum01} Comparison of the density estimates of the distributions of the parameters using an ABC approximation
with $\epsilon$ as the $0.1\%$ quantile on the autocovariance distances {\em (in blue)} and the \cite{beaumont:zhang:balding:2002} correction
{\em (in red)}. The red and blue curves are confounded for the parameter $\theta_2$.}
\end{figure}

\begin{figure}[hbtp]
\centering\includegraphics[width=.5\textwidth]{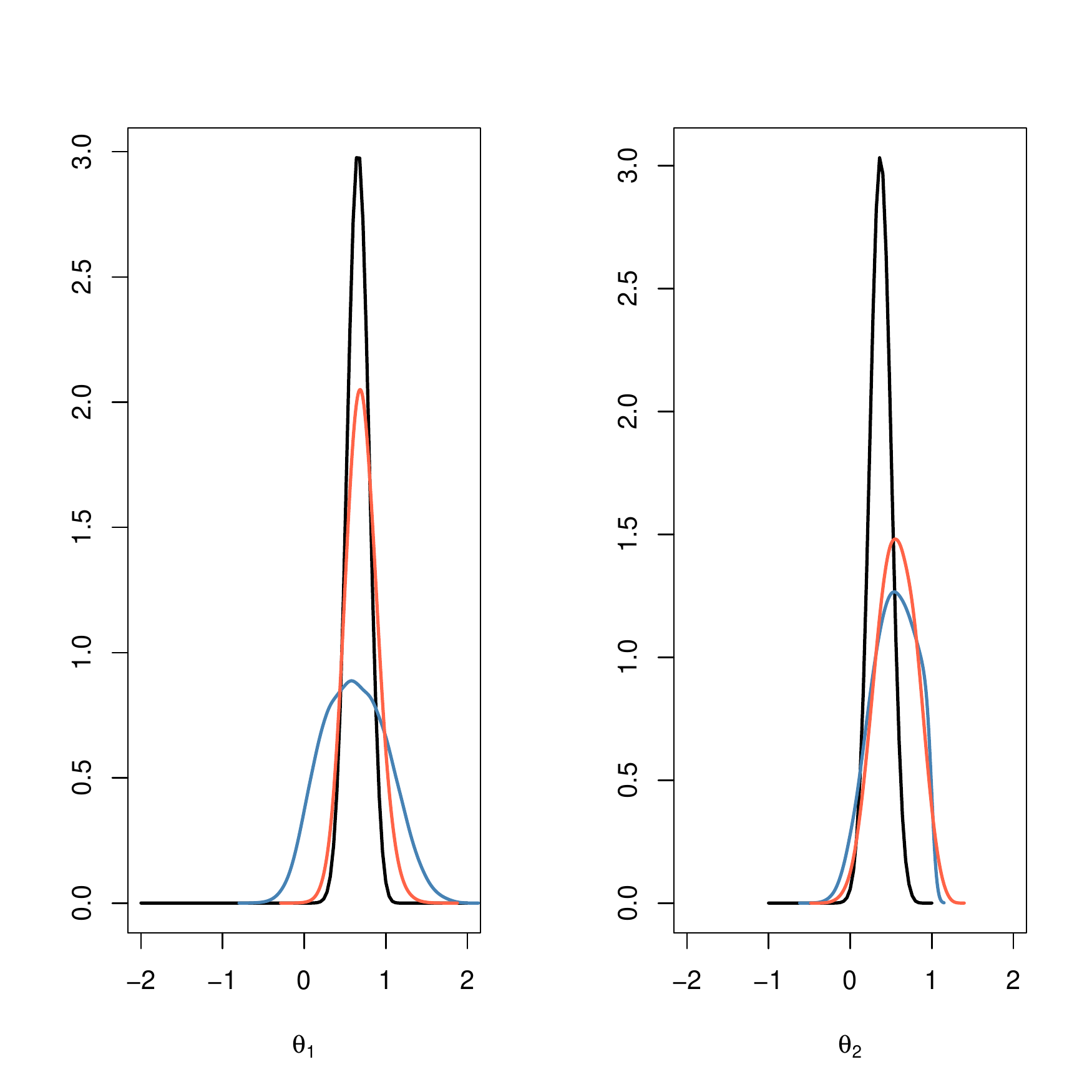}
\caption{\label{fig:BeAum20} Comparison of the approximate distributions of the parameters using an ABC approximation
with $\epsilon$ as the $20\%$ quantile on the autocovariance distance {\em (in blue)} and the \cite{beaumont:zhang:balding:2002} 
correction {\em (in red)}.}
\end{figure}

\begin{ex}We implement this correction of \cite{beaumont:zhang:balding:2002} in the MA$(2)$ model,  again
using the first two autocovariances as summary statistic $\eta(\bz)$, and we apply a non-parametric local regression 
based on the Epanechnikov kernel. We keep $\delta$ equal to the value of the tolerance $\epsilon$ used in the regular ABC 
scheme. Figures \ref{fig:BeAum01} and \ref{fig:BeAum20} summarise the results. When using a $0.1\%$ quantile, the two density
estimates are identical in the case of the parameter $\theta_2$. The post-processed density estimate of $\theta_1$ is closer to the
true posterior. When using a $20\%$ quantile, the impact of the local regression is more spectacular. We  recover  results close to those
obtained with the $0.1\%$ quantile. This exhibits the point that local regression strongly attenuates the impact of the truncation
brought by $\epsilon$.
\end{ex}


\paragraph{\bf Nonlinear regression}
\cite{blum:francois:2010} propose a generalisation of \citeauthor{beaumont:zhang:balding:2002}'s (2002) ABC
post-processing where the local linear regression of the parameter $\btheta$ on the summary statistics $\eta(\bz)$ is
replaced by a nonlinear regression with heteroskedasticity. In this new approach, the nonlinear mean and variance are
estimated by a neural net with one hidden layer, using the R package {\sf nnet} \citep{cran}. The result is interesting
in that it seems to allow for the inclusion of more or even all the simulated pairs $(\btheta,\bz)$, compared with
Beaumont et al. (2002). This is somehow to be expected since the nonlinear fit adapts differently to different parts of
the space.  Therefore, weighting simulated $(\btheta,\bz)$'s by a kernel $K_\delta(\bz-\by)$ is not very relevant and it is thus
not surprising that the bandwith $\delta$ is not influential, in contrast with  basic ABC and even Beaumont et al.
(2002) where $\delta$ has a different meaning. The non-parametric perspective adopted in the paper is nonetheless of the
highest importance, as it proves the most fruitful approach to the interpretation of ABC methods. In connection with
this paper, \cite{blum:2010} provides a good review of the non-parametric handling of ABC techniques.  The true
difficulty with the non-parametric perspective lies with the curse of dimensionality. This issue might be addressed by mixing dimension reduction with
recycling by shrinking as in Beaumont et al. (2002).

\paragraph{\bf Inverse regression}
\cite{leuenberger:wegmann:excoffier:2010} also relate to the local regression ideas in Beaumont et al. (2002).
As in the earlier work by \cite{wilkinson:2008}, the approximation to the
distribution of the parameters given the observed summary statistics is central to the paper. In opposition to
\cite{beaumont:zhang:balding:2002}, there is no clear shrinkage for summary statistics that are far away from the
observed summary statistics: all accepted parameters are weighted similarly in the Gaussian linear approximation to the
truncated prior. The other difference with \cite{beaumont:zhang:balding:2002} is that the authors model $\bz$ given
$\btheta$ rather than $\btheta$ given $\bz$, in an inverse regression perspective, followed by a sort of Laplace
approximation reminding \cite{rue:martino:chopin:2009}.

\section{ABC and model choice}\label{RoxyMusic} 

\subsection{Bayesian model choice}
Model choice is one particular aspect of Bayesian analysis that involves computational complexity, if only
because several models are considered simultaneously \citep[see, e.g.,][]{robert:2001,marin:robert:2010}. In addition to the
parameters of each model, the inference considers the model index $\mathcal{M}$, which is associated with its
own prior distribution  $\pi(\mathcal{M}=m)$ ($m=1,\ldots,M$) as well as a prior distribution on the parameters conditional on
the value $m$ of the model index, $\pi_m(\btheta_m)$, defined on the parameter space $\Theta_m$.  The choice
between these models is then driven by the posterior distribution of $\mathcal{M}$, a challenging computational target
where ABC brings a straightforward solution. Indeed, once $\mathcal{M}$ is incorporated within the parameters,
the ABC approximation to the posterior follows from the same principles as regular ABC, as shown by the
following pseudo-code, where $\feta(\bz)=(\eta_1(\bz),\ldots,\eta_M(\bz))$ is the concatenation of the summary
statistics used for all models (with elimination of duplicates).

\begin{algorithm}\caption{Likelihood-free model choice sampler (ABC-MC) \label{algo:ABCMoC}} 
\begin{algorithmic} 
\FOR {$i=1$ to $N$} 
\REPEAT 
\STATE Generate $m$ from the prior $\pi(\mathcal{M}=m)$ 
\STATE Generate $\btheta_{m}$ from the prior $\pi_{m}(\btheta_m)$ 
\STATE Generate $\bz$ from the model $f_{m}(\bz|\btheta_{m})$ 
\UNTIL {$\rho\{\feta(\bz),\feta(\by)\}<\epsilon$} 
\STATE Set $m^{(i)}=m$ and $\btheta^{(i)}=\btheta_m$ 
\ENDFOR 
\end{algorithmic} 
\end{algorithm} 
The ABC estimate of the posterior probability $\pi(\mathcal{M}=m|\by)$ is then the acceptance frequency from model
$m$, namely
\begin{equation*}
\dfrac{1}{N}\,\sum_{i=1}^N \mathbb{I}_{m^{(i)}=m}\,. 
\end{equation*}
This also corresponds to the proportion of simulated datasets that are closer to the data $\by$ than the tolerance $\epsilon$.  
\cite{cornuet:santos:beaumont:etal:2008} follow the rationale that
led to the local linear regression in \cite{beaumont:zhang:balding:2002} and rely on a weighted polychotomous
logistic regression to estimate $\pi(\mathcal{M}=m|\by)$. 
This modeling clearly brings some further stability to the above estimate of $\pi(\mathcal{M}=m|\by)$ and
is implemented in the DIYABC software described in  \cite{cornuet:santos:beaumont:etal:2008} .
 
\begin{figure}[hbtp]
\centering
\includegraphics[width=.5\textwidth,height=.45\textwidth]{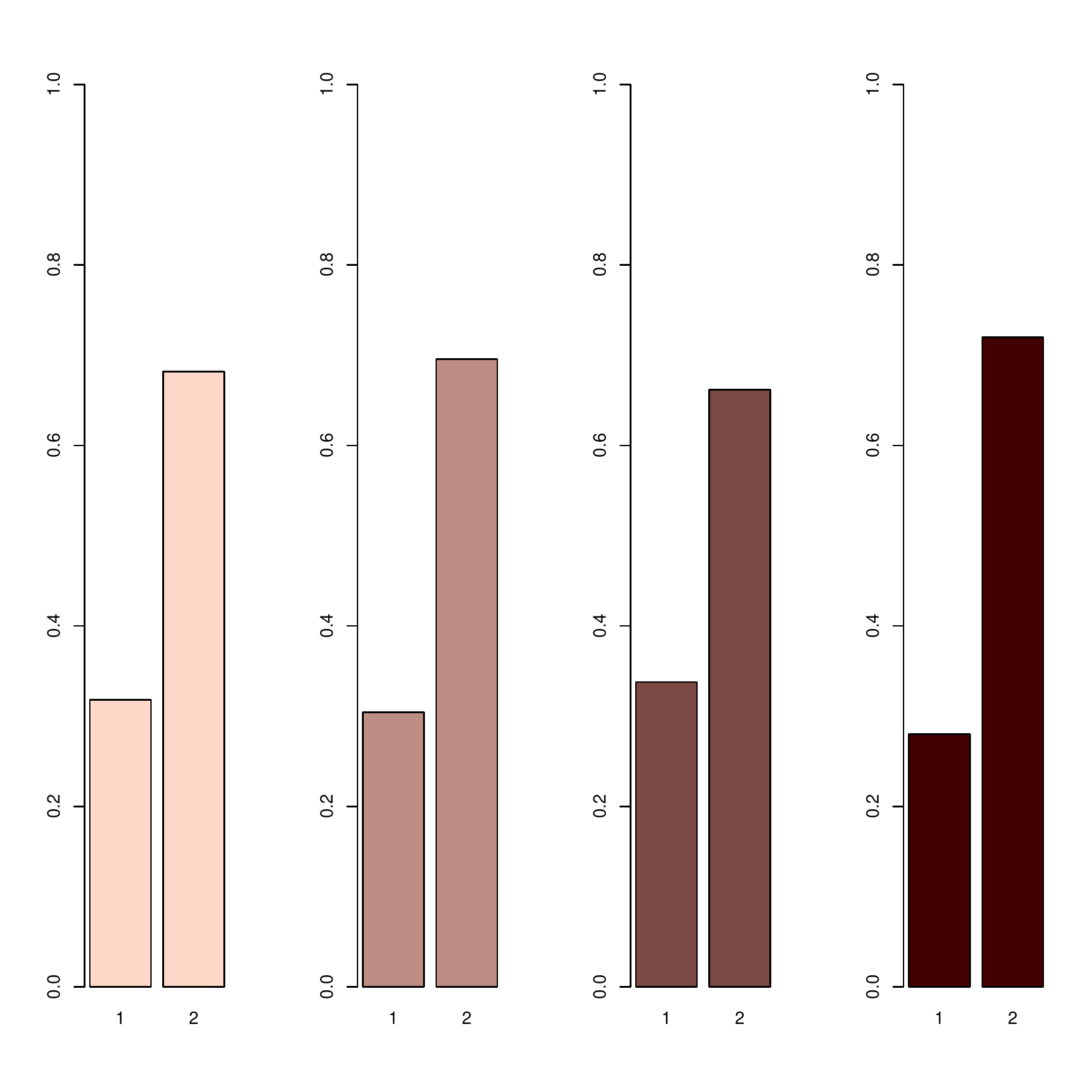}
\caption{\label{fig:Moc2} Boxplots of the evolution [against $\epsilon$] of ABC approximations to the Bayes factor.
The representation is made in terms of frequencies of
visits to [accepted proposals from] models MA$(1)$ \textit{(left)} and MA$(2)$ \textit{(right)} during an ABC simulation
when $\epsilon$ corresponds to the $10,1,0.1,0.01\%$ quantiles on the simulated autocovariance distances. The data are the same
as in Figure \ref{fig:BeAum20}. The true Bayes factor $B_{21}$ is equal to $17.71$, corresponding to
posterior probabilities of $0.05$ and $0.95$ for the MA(1) and MA(2) models respectively.}
\end{figure}

\begin{figure}[hbtp]
\centering
\includegraphics[width=.5\textwidth,height=.45\textwidth]{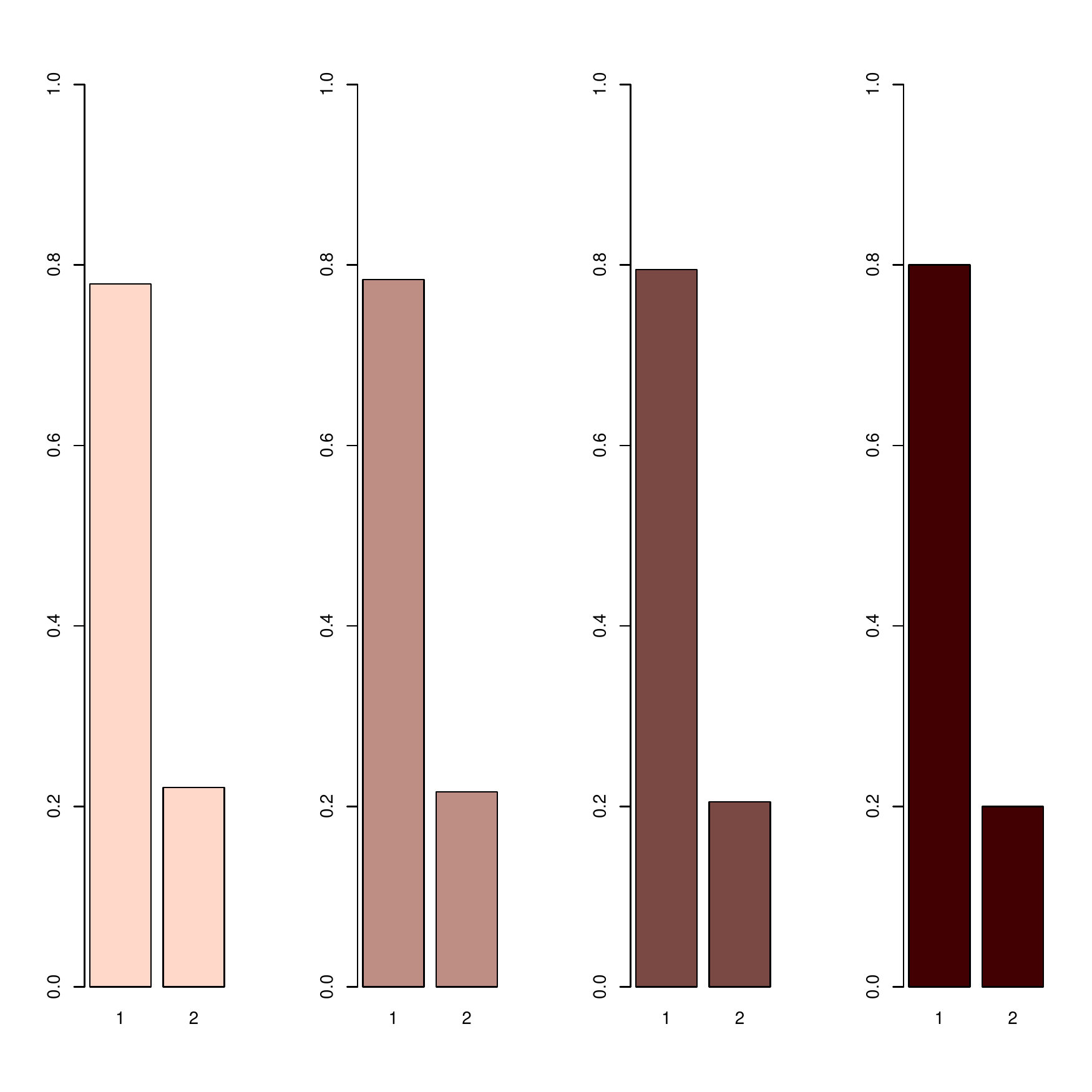}
\caption{\label{fig:Moc1} Boxplots of evolution of Bayes factor approximations in terms of frequencies of
visits to models MA$(1)$ \textit{(left)} and MA$(2)$ \textit{(right)} using an ABC approximation
with $10,1,.1,.01\%$ quantiles on the autocovariance distance as $\epsilon$. The dataset is a sample of $50$
points from a MA$(1)$ model with $\theta_1=0.6$. The true Bayes factor $B_{21}$ is equal to $.004$ corresponding to
posterior probabilities of $0.996$ and $0.004$ for the MA(1) and MA(2) models respectively.}
\end{figure}

\begin{ex} Returning once again to our benchmark MA$(2)$ model, we compare the computation of the model posterior probabilities
based on an ABC sample (acceptance frequency  within each model) with the true value of the Bayes factor, which was obtained by numerical integration. The dataset used in the experiment
is a time-series simulated  and we wish to choose between two models: an MA$(2)$ or an MA$(1)$ model. 
Figure \ref{fig:Moc2} shows our estimates for data simulated from ar MA$(2)$ model. The weight of the MA$(2)$ model  increases slightly as $\epsilon$ decreases. However, even for the quantile at
$0.01\%$ the estimated posterior probability for the MA(2) model is equal to $0.72$ which is far from the true value $0.95$. 
Figure \ref{fig:Moc1} shows a similar phenomenon for data simulated from an MA$(1)$ model. 
\end{ex}

The discrepancy in the above example shows the limitations of the ABC approximation of Bayes factors exposed in
\citet{robert:cornuet:marin:pillai:2011}. While we could expect to obtain a better approximation with a massive
computational effort, it may be that the use of different summary statistics for different models prevents us
from converging to the true value. In other words, the concatenation of sufficient statistics for individual
models does not always constitute a sufficient statistic for model choice, as discussed in the next paragraph.

\subsection{The case of Gibbs random fields}
\cite{grelaud:marin:robert:rodolphe:tally:2009} show that, for Gibbs random fields and in particular for Potts models,
where the goal is to compare several neighbourhood structures, the computation of the posterior probabilities of the
models under competition can be operated by likelihood-free simulation techniques. We recall first that Gibbs
random fields are probabilistic models associated with the likelihood function 
\begin{equation*}\label{Gee}
\ell(\btheta|\by)=\dfrac{1}{Z_{\btheta}}\exp\{\btheta^\text{T}\eta(\by)\}\,, 
\end{equation*} 
where $\by$ is a vector of dimension $n$ taking values over a finite set $\mathcal{X}$ (possibly a lattice), $\eta(\cdot)$ is the potential function
defining the random field, taking values in $\mathbb{R}^p$, $\btheta\in\mathbb{R}^p$ is the associated parameter, and
$Z_{\btheta}$ is the corresponding normalising constant. A special but important case of Gibbs random fields is
associated with a neighbourhood structure denoted by $i \sim i'$ (meaning that $i$ and $i'$ are neighbours), in that
\begin{equation*}
\eta(\by)=\sum_{i' \sim i} \mathbb{I}_{\{y_i=y_{i'}\}}\,,
\end{equation*}
where $\sum_{i' \sim i}$ indicates that the summation is  over all the pairs of neighbours.
In that case, $\btheta$ is a scalar. 

The central property ensuring an ABC resolution for Gibbs random fields is that, due to their exponential
family structure, there exists a sufficient statistic vector that runs across models and which allows for an
exact ($\epsilon=0$) simulation from the posterior probabilities of the models. Indeed, model choice involves
$M$ Gibbs random fields in competition; each field is  associated with a potential function $\eta_m$ $(1\le
m\le M)$, \textit{i.e.}~with the corresponding likelihood 
\begin{equation*}
\ell_m(\btheta_m|\by)=\exp \left\{ \btheta_m^\text{T} \eta_m(\by) \right\} \big/ Z_{\btheta_m,m}\,,  
\end{equation*}
where
$\btheta_m\in\Theta_m$ and $Z_{\btheta_m,m}$ is the unknown normalising constant.  From a Bayesian perspective,
considering an extended parameter space $\Theta=\cup_{m=1}^{M}\{m\}\times\Theta_{m}$ that includes the model index
$\mathcal{M}$, the computational target is thus the model posterior probability 
\begin{equation*}
\pi(\mathcal{M}=m|\by)\propto\int_{\Theta_m} \ell_m(\btheta_m|\by) \pi_m(\btheta_m) \,\text{d}\btheta_m\,\pi(\mathcal{M}=m)\,, 
\end{equation*}
\textit{i.e.}~the marginal in $\mathcal{M}$ of the posterior distribution on $(\mathcal{M},\btheta_1,\ldots,\btheta_M)$ given
$\by$.  Each model has its own sufficient statistic $\eta_m(\cdot)$. Then, for each {\em individual} model, the
vector of statistics $\feta(\cdot)=(\eta_1(\cdot),\ldots,\eta_M(\cdot))$ is clearly sufficient. However
\cite{grelaud:marin:robert:rodolphe:tally:2009} exposed the fact that $\feta$ is also sufficient for the {\em joint} parameter
$(\mathcal{M},\btheta_1,\ldots,\btheta_M)$. 

That the concatenation of the sufficient statistics of each model is also a sufficient statistic for the joint
parameter across models is clearly a property that is specific to exponential families. As shown by
\citet{didelot:everitt:johansen:lawson:2011}, ABC-based model choice can process exponential families by
creating inter-model sufficient statistics that incorporate the intra-model sufficient statistics as well as
possibly the dominating measures for all models. The Gibbs random field above is a specific case of this
sufficiency. However, outside exponential families, the possibility of creating a sufficient statistic of a
dimension that is much lower than the dimension of the data is impossible, as explained in
\citet{robert:cornuet:marin:pillai:2011}.

\subsection{General issues}
\cite{toni:etal:2009} and \cite{toni:stumpf:2010} review ABC-based model choice, inclusive of the above Gibbs random
field example. The authors study in particular the consequences of implementing a sequential algorithm like ABC-PMC in this
set-up. The ABC algorithm is modified to incorporate the model index, resorting to the previous assessment of
$\pi(\mathcal{M}=m|\by)$ to propose the model indices of the next population. The importance sampling features of this
setting imply that the posterior probability can be estimated from the importance weights. However, the adaptivity at
the core of ABC-PMC and ABC-SMC implies adapting an approximation kernel for each model.
As most other perspectives on ABC, \cite{toni:stumpf:2010} do not question the role of the ABC distance in
model choice settings.  The Bayes factors are observed to be sensitive to the choice of the prior distributions, of the
tolerance levels, and to the variances of the kernels $K_t$ (see Section \ref{Scorpion}), a dependence that should not
occur, since this is a simulation parameter that is unrelated with the statistical problem.

It is worth pointing out the remark made by \cite{leuenberger:wegmann:excoffier:2010} about model choice and the use
of the approximation of the normalising constant resulting from the modelling to get to the marginal likelihood and the
computation of the Bayes factor. This relates to earlier comments in the literature about the ABC acceptance rate
approximating the marginal and a recent paper by \cite{bartolucci:scaccia:mira:2006} studying
ways of computing marginal probabilities by Rao--Black\-wel\-li\-sing reversible jump acceptance probabilities.
\cite{grelaud:marin:robert:rodolphe:tally:2009} also make the most of this ABC feature for Ising models, since an
exact ABC (corresponding to $\epsilon=0$) algorithm is then available for model selection.

A (minor) Bayesian issue mentioned by \cite{ratmann:andrieu:wiujf:richardson:2009} is the fact that both $\btheta$ and
$\epsilon$ are taken to be the same across models. In a classical Bayesian perspective, modulo the
reparameterisation, $\btheta$ cannot be entirely different from one model to the next, but using the same prior on
$\epsilon$ over all models under comparison is more of an issue. The paper also considers the impact of testing for the
adequacy of a model as testing for the hypothesis $H_0:\,\epsilon=0$, an interesting if controversial stance, since even
when the model fits, $\epsilon$ necessarily varies around zero. 

At this stage, the most perplexing feature of ABC model choice is the lack of convergence guarantees. As
exposed in \citet{robert:cornuet:marin:pillai:2011}, most settings where ABC model choice is implemented do not
allow for inter-model sufficiency in the selection of the summary statistics, because some models are not within
exponential families and because using the whole data is too demanding. As shown by the MA example above. this
lack of sufficiency may be quite detrimental to the quality of the ABC approximation of the Bayes factors.
There is therefore currently no theoretical support for the use of ABC approximations of Bayes factors and
posterior model probabilities, and we thus advise for more empirical assessments in the spirit of
\citet{ratmann:andrieu:wiujf:richardson:2009} that evaluate the model fit within each model without concluding
by exact figures of the probabilities of the different models.
 
\section{Discussion}

Approximate Bayesian Computation allows inference from a wide class of models which would otherwise be unavailable. As such, it has spawned interest in both theoretical issues and applications. Recent advances regarding the calibration of the method lead to an approximation that is good enough to be highly useful in many situations. The efficiency of the method can be greatly improved with sequential techniques and post-processing regression on the output.

Nonetheless, ABC is not a silver bullet. In the current state of the art, it can only be used for model choice in a limited range of models. Future advances  must at the same time expand further the tools to make ABC useful in a wider class of models, extend pre- and post-processing methods to control the approximation, and establish more clearly in which cases ABC reaches its limitations.

\newpage

ABC methods are currently under an intense scrutiny by both statisticians and practitioners, hence the object of an unparalleled development. While this rapid development provides answers to some interrogations from the statistical community about the validity of the approach and from the practitioners about a higher efficiency of the method, some issues remain unsolved, among which:

\begin{itemize}
\item the convergence results obtained so far are unpractical in that they require either the tolerance to go to zero or the sample size to go to infinity.
Obtaining exact error bounds for positive tolerances and finite sample sizes would bring a strong improvement in both the implementation of
the method and in the assessment of its worth.

\item even though ABC is often presented as a converging method that approximates Bayesian inference, it can also be perceived as an inference technique per
se and hence analysed in its own right. Connections with indirect inference have already been drawn, however the fine asymptotic analysis of ABC would be most useful
to derive. Moreover, it could indirectly provide indications about the optimal calibration of the algorithm.

\item in connection with the above, the connection of ABC-based inference with other approximative methods like variational Bayes inference is so far unexplored.
Comparing and interbreeding those different methods should become a research focus as well.

\item the construction and selection of the summary statistics is so far highly empirical. An automated approach based on the principles of data analysis and
approximate sufficiency would be much more attractive and convincing, especially in non-standard and complex settings.

\item the debate about ABC-based model choice is so far inconclusive in that we cannot guarantee the validity of the approximation, while considering that
a ``large enough" collection of summary statistics provides an acceptable level of approximation. Evaluating the discrepancy by exploratory methods like
the bootstrap would shed a much more satisfactory light on this issue.

\item the method necessarily faces limitations imposed by large datasets or complex models, in that simulating pseudo-data may itself become an impossible task.
Dimension-reducing technique that would simulate directly the summary statistics will quickly become necessary.
\end{itemize}

\section*{Acknowledgements}

The authors are grateful to J.-M. Cornuet for bringing the problem to their attention and for a highly enjoyable and fruitful collaboration over the past years. 
Part of this work was conducted while the third author was visiting the Department of
Statistics at the Wharton Business School of the University of Pennsylvania, to whom he is most grateful for its support.
The authors are also grateful to J.-L. Fouley for several interesting discussions.

\bibliographystyle{apalike}
\bibliography{MPRR11.bib}

\end{document}